\documentclass[prl,twocolumn,superscriptaddress]{revtex4}

\usepackage{amsbsy}

\newcommand{\COMMENT}[1]{}

\begin{document}

\title{Fully Distrustful Quantum Cryptography}

\author{J. Silman}
\affiliation{Laboratoire d'Information Quantique, Universit{\' e}
Libre de Bruxelles, 1050 Bruxelles,
Belgium}
\author{A. Chailloux}
\affiliation{LIAFA, Univ. Paris 7, F-75205 Paris, France; and Univ. Paris-Sud, 91405 Orsay, France}
\author{N. Aharon}
\affiliation{School of Physics and Astronomy, Tel-Aviv University, Tel-Aviv 69978, Israel}
\author{I. Kerenidis}
\affiliation{LIAFA, Univ. Paris 7 - CNRS; F-75205 Paris, France}
\affiliation{Centre for Quantum Technologies, National University of Singapore, Singapore 117543}
\author{S. Pironio}
\author{S. Massar}
\affiliation{Laboratoire d'Information Quantique, Universit{\' e}
Libre de Bruxelles, 1050 Bruxelles,
Belgium}

\begin{abstract}
In the distrustful quantum cryptography model the different parties
have conflicting interests and do not trust one another. Nevertheless,
they trust the quantum devices in their labs. The aim of the
device-independent approach to cryptography is to do away with the
necessity of making this assumption, and, consequently, significantly
increase security. In this paper we enquire whether the scope of the
device-independent approach can be extended to the distrustful
cryptography model, thereby rendering it `fully' distrustful. We
answer this question in the affirmative by presenting a
device-independent (imperfect) bit-commitment protocol, which we then
use to construct a device-independent coin flipping protocol.

\end{abstract}


\maketitle

{\em Introduction} -- A quantum protocol is said to be device-independent if the reliability of its 
implementation can be guaranteed without
making any assumptions regarding the internal workings of
the underlying apparatus. 
The key idea is that the certification of a sufficient amount of nonlocality ensures that the underlying systems are quantum and entangled. 
By dispensing with the
(mathematically convenient but physically untestable) notion of a Hilbert
space of a fixed dimension, the {device-independent} approach does away with many cheating mechanisms and modes of failure,
such as, for example, those exploited in \cite{lo,makarov}. 
In fact, a {device-independent} protocol, in principle, remains secure even if the devices were fabricated
by an adversary. So far, {device-independent} protocols have been proposed for quantum
key-distribution \cite{Acin et al. 4,Acin et al., Mayers 0,Barrett et al.},
random number generation \cite{Colbeck,Pironio et al. II}, state estimation \cite{Bardyn et al.}, and the self-testing of quantum computers \cite{Magniez et al.}.

In many everyday
scenarios (e.g. the use of credit cards on the internet, secure identification, digital signatures), we need to ensure
security not only against an eavesdropper, but crucially against malicious parties partaking in the protocol, i.e. when Alice and
Bob do not trust each other. Many important results in quantum cryptography are related to the fundamental
primitives in this setting: While, on the one hand, quantum weak coin flipping with arbitrarily small bias is
possible \cite{Mochon III}, arbitrarily concealing and binding quantum bit-commitment is impossible
\cite{Lo, Mayers 2, Werner}. However, less secure but non-trivial {bit-commitment} has been shown to be possible with trusted devices~\cite{Spekkens & Rudolph}.

It is not a priori clear, whether the scope of the {device-independent} approach can be extended to cover 
cryptographic problems with distrustful parties. In particular, this setting presents us with a novel challenge:
Whereas in {device-independent} quantum key-distribution Alice and Bob will cooperate to estimate the amount of nonlocality present, for protocols in the distrustful cryptography model, honest parties can rely only on themselves. In this paper we show that protocols in this model are indeed amenable to a {device-independent} formulation.
As our aim is to provide a proof of concept, we concentrate on one of the simplest, yet most fundamental, primitives in this model, {bit-commitment.
We present a {device-independent} {bit-commitment} protocol, wherein after the commit phase Alice cannot control the value of the bit she wishes to reveal with probability greater than $\cos^2\left(\frac{\pi}{8}\right)\simeq 0.854$ and Bob cannot learn its value prior to the reveal phase with probability greater than $\frac{3}{4}$. We then use this protocol to construct a {device-independent} {coin flipping} protocol with bias $\lesssim 0.336$.

{ \em Bit-commitment} -- A {bit-commitment} protocol consists of two phases.
In the commit phase, Alice interacts with Bob in order to commit to
a bit. In the reveal phase, Alice reveals the value of the bit, possibly followed by some test
that each party carries out to ensure that the other party has
not cheated. In the time between the two phases, which may be of any duration, no actions are taken.
The security of a protocol is always analyzed under
the assumption that one of the parties is honest. 
We designate by $P_\mathrm{cont}$, 
the maximum of the average of the probabilities with which Alice can reveal either value of the bit without being caught cheating, 
and by $P_\mathrm{gain}$ the maximum probability that dishonest Bob learns the value of bit before the reveal phase without being discovered, where these quantities are maximized over the set of possible cheating strategies available to Alice and Bob.
The quantities
$\epsilon_{\mathrm{cont}}=P_{\mathrm{cont}}-\frac{1}{2}$ and $\epsilon_{\mathrm{gain}}=P_{\mathrm{gain}}-\frac{1}{2}$ are termed
`Alice's control' and `Bob's information gain'. A protocol with arbitrarily small $\epsilon_{\mathrm{cont}}$ is called arbitrarily binding,
while a protocol with arbitrarily small $\epsilon_{\mathrm{gain}}$ is called arbitrarily concealing. As already mentioned, quantum mechanics
does not allow for a protocol to be both arbitrarily binding and concealing
at the same time.
In fact, for a `fair' protocol, in the sense that $\epsilon_{\mathrm{cont}}=\epsilon_{\mathrm{gain}}$,
$\epsilon_{\mathrm{cont}}$ is bounded from below by $0.207$ \cite{footnote}. The
best known protocol gives $\epsilon_{\mathrm{cont}}=\frac{1}{4}$ \cite{Spekkens & Rudolph}.
In contrast, in any classical protocol either Alice or Bob can cheat perfectly ($\epsilon_{\mathrm{cont}}=\frac{1}{2}$).

{\em Device-independence} -- In our device-independent formulation, we assume that each honest party has one or several devices which are viewed as `black boxes'. Each box allows for a classical input $s_{i}\in\left\{0,1\right\} $, and produces a classical output $r_{i}\in\left\{0,1\right\}$ (the index $i$ designates the box). We make the assumption that the probabilities of the outputs given the inputs for an honest party can be expressed as 
$P(\mathbf{r}|\mathbf{s})=\mathrm{Tr}\bigl(\rho\bigotimes_{i}\Pi_{r_{i}\mid s_{i}}\bigr)$, where 
$\rho$ is some joint quantum state and $\Pi_{r_{i}\mid s_{i}}$
is a POVM element corresponding to inputting $s_{i}$ in box $i$
and obtaining the outcome $r_{i}$. Apart from this constraint we impose no restrictions on the
boxes' behavior. In particular, we allow a dishonest party to choose
the state $\rho$ (which she can entangle with her system) and the POVM elements $\Pi_{r_{i}\mid s_{i}}$ for the other party's boxes.

The above assumption amounts to the most general modeling of boxes that (i) satisfy the laws of quantum theory, and (ii) are such that the physical process yielding the output $r_i$ in box $i$ depends solely on the input $s_i$, i.e. the boxes cannot communicate with one another. It is also implicit in our analysis that no unwanted information can enter or exit an honest party's laboratory. 
 In a `fully' distrustful setting, where the devices too cannot be trusted, these conditions can be satisfied by shielding the boxes. In particular, it is not necessary to carry out measurements in space-like separated locations to guarantee (ii), as in fundamental tests of nonlocality (see \cite{Pironio et al. I,Pironio et al. II}).
This observation is important because relativistic causality is by itself sufficient for perfect {bit-commitment} and {coin flipping} \cite{Kent, Kent 2}. Hence, the fact that we do not rely on space-like measurements makes
the conceptual implications of our work clearer and the quantum origin of the security evident.

{\em The protocol} -- Our protocol is based on the Greenberger-Horne-Zeilinger (GHZ) paradox \cite{GHZ, Mermin}. We consider three boxes $A$, $B$, and $C$ with binary inputs, $s_A$, $s_B$ and $s_C$, and outputs $r_A$, $r_B$ and $r_C$, respectively. The GHZ paradox consists of the fact that if the inputs satisfy $s_A\oplus s_B\oplus s_C=1$, we can always have the outputs satisfy $r_A\oplus r_B\oplus r_C=s_A s_B s_C\oplus1$. This relation can be guaranteed if the three boxes implement measurements on a three-qubit GHZ state $\frac{1}{\sqrt{2}}(|000\rangle+|111\rangle)$, where $s_i=0$ $(1)$ corresponds to measuring $\sigma_y$ ($\sigma_x$).
In contrast, for local boxes this relation can only be satisfied with $\frac{3}{4}$ probability at most. 
The nonlocal and pseudo-telepathic nature of the GHZ paradox -- the non-occurrence of certain input-output pairs that would necessarily
occur in any local theory -- are key, both to ensure that when both parties are honest the protocol does
not abort, and to ensure that a dishonest party always has a non-zero
probability of being caught cheating.

The protocol runs as follows. Alice has a box, $A$, and Bob has a pair of boxes, $B$ and $C$. The three boxes are supposed to satisfy the GHZ paradox. {\em Commit phase}: Alice inputs into her box the value of the bit she wishes to commit to. Denote the input and output of her box by $s_A$ and $r_A$. She then selects a classical bit $a$ uniformly at random. If $a=0$ ($a=1$), she sends Bob a classical bit $c=r_A$ ($c=r_A \oplus s_A$) as her commitment. {\em Reveal phase}: Alice sends Bob $s_A$ and $r_A$. Bob first checks whether $c=r_A$ or $c=r_A\oplus s_A$. He then randomly chooses a pair of inputs $s_B$ and $s_C$, satisfying $s_B\oplus s_C=1\oplus s_A$, inputs them into his two boxes and checks that the GHZ paradox is satisfied. 
If any of these tests fails then he aborts. Note that if the parties are honest (and the boxes satisfy the GHZ paradox), then the protocol never aborts.

{\em Alice's control} -- We consider the worst-case scenario, wherein (dishonest) Alice prepares (honest) Bob's boxes in any state she wants, possibly entangled with her own ancillary systems. Since the commit phase consists of Alice sending a classical bit $c$ as a token of her commitment, without receiving any information from Bob, with no loss of generality we may assume that Alice decides on the value of $c$ beforehand, and accordingly prepares Bob's boxes to maximize her control. Furthermore, since Alice's winning probability is invariant under the relabeling, $c\to c\oplus1$, $r_A\to r_A\oplus1$,
$r_B\to r_B\oplus1$, no value of $c$ is preferable, and we assume that she sends $c=0$.

Suppose now that Alice wishes to reveal $0$ (i.e. she
sends $s_A=0$). She will then carry out some operation on her systems in order to decide the value of $r_A$ to be sent. Bob will first check whether $r_A=0$ or $r_A\oplus s_A=0$,
and since $s_A=0$ it follows that Alice must send $r_A=0$. Subsequently,
Bob finds that the GHZ paradox is satisfied whenever $r_B \neq r_C$
for a choice of inputs such that $s_B \neq s_C$. Switching to a more compact notation in which $y_i=\left(-1\right)^{r_i}$ ($x_i=\left(-1\right)^{r_i}$) designates the output corresponding to $s_i=0$ ($s_i=1$), Alice's cheating
probability in this case equals $\frac{1}{2}\left[P\left(y_B x_C=-1\right)+P\left(x_B y_C=-1\right)\right]$.
 On the other hand, suppose that Alice wishes to reveal $1$. Then,
$r_A$ may take on any value (since Bob knows that in this case $r_A=0$ or $r_A\oplus1=0$),
and hence, the only relevant test is the satisfaction of the GHZ paradox, i.e.
whether $r_B\oplus r_C=s_B s_C\oplus1\oplus r_A$ for a choice
of inputs such that $s_B=s_C$. Alice's cheating probability
then equals $\frac{1}{2}\left[P\left(y_A y_B y_C=-1\right)+P\left(x_A x_B x_C=1\right)\right]$.
Hence, Alice's optimal cheating probability is obtained by maximizing over 
\begin{eqnarray}
  &  & \frac{1}{4}\bigl[P\left(y_{B}x_{C}=-1\right)+P\left(x_{B}y_{C}=-1\right)\bigr.\nonumber \\
& & \bigl.+P\left(x_{A}y_{B}y_{C}=-1\right)+P\left(x_{A}x_{B}x_{C}=1\right)\bigr]\,
\end{eqnarray}
since we consider the average probability that Alice can
reveal 0 and 1.
 As this expression involves only a single measurement setting for Alice's box, it admits a local description, implying that the maximum is obtained when
Alice's box is deterministic. We see that in both cases (i.e. $x_{A}=\pm1$),
the problem reduces to maximizing the Clauser-Horne-Shimony-Holt (CHSH) inequality \cite{CHSH}, so that $P_{\mathrm{cont}}=\cos^{2}\left(\frac{\pi}{8}\right)\simeq0.854$.

{\em Bob's information gain} --
Bob's most general strategy consists of sending Alice a box entangled
with some ancillary system in his possession. Depending on the value
of $c$ he receives from Alice (which is uniformly random since Alice is honest), Bob carries out one of a pair of two-outcome
measurements on his system. We denote Bob's binary input and output
by $m_B$ and $g_B$, where $m_B=0$ ($m_B=1$)
corresponds to the measurement he carries out when Alice sends $c=0$
($c=1$), and $g_B=0$ ($g_B=1$) corresponds to his guessing that Alice has committed to $0$ ($1$).
Bob's information gain is 
\begin{eqnarray}
\lefteqn{P_{\mathrm{gain}}}\nonumber\\
 & = & 
\sum_{s_A,r_A,\,a} P(s_A,r_A,\,a)P\left(g_B=s_A\mid m_B=r_A\oplus (s_A\cdot a)\right) \nonumber\\
& = & \quad\quad \frac{1}{4}\sum_{s_A,\,r_A=0,\,1}P\left(r_A\mid s_A\right)\bigl[P\left(g_B=s_A\mid m_B=r_A\right)\bigr.\nonumber\\
& & \bigl.+P\left(g_B=s_A\mid m_B=r_A\oplus s_A\right)\bigr]\nonumber \\
 & = & \frac{1}{4}\sum_{s_A,\,r_A=0,\,1}\bigl[P\left(r_A,\, g_B=s_A\mid s_A,\, m_B=r_A\right)\bigr. \nonumber\\ 
& &  \quad\quad\bigl.+P\left(r_A,\, g_B=s_A\mid s_A,\, m_B=r_A\oplus s_A\right)\bigr]
 \,.\end{eqnarray} 
Using the fact that 
$P(k,\,0|0,\,k)+P(0,\,1|1,\,k)+P(1,\,1|1,\,k)\leq 1$
and
$P\left(0,\,0\mid0,\,0\right)+P\left(1,\,0\mid0,\,1\right)\leq1$,
which follow
from no-signaling (i.e. $\sum_{l=0,\,1}P\left(i_A,\,i_B\mid j_A,\,j_B\right)=P\left(i_A\mid j_A\right)$ and the same relation with $A\leftrightarrow B$) and normalization,
we obtain that  $P_{\mathrm{gain}}=\frac{3}{4}$.

{\em Optimal cheating strategies} -- Both Alice and Bob have a number of simple optimal cheating strategies available to them. Interestingly, both can optimally cheat using a three-qubit GHZ state and having the measurements of the honest party correspond to the measurement of $\sigma_y$ and $\sigma_x$ axes (corresponding to inputting $0$ and $1$), as in the GHZ paradox described above. This implies that the device-dependent version of our protocol, in which (honest) Alice and Bob share a GHZ state and measure $\sigma_{y}$ and $\sigma_{x}$ (recall that in the {device-dependent} setting an honest party can trust its measurement devices), does not afford more security. Our protocol has thus the curious property that its {device-dependent} version is essentially {device-independent}, in the sense that its security is not compromised in the event that an honest party cannot trust its measurement devices.

Using the GHZ state, dishonest Alice's strategy consists of measuring the polarization of her qubit along the axis $\hat{n}=\frac{1}{\sqrt{2}}\left(\hat{x}+\hat{y}\right)$. If she obtains $0$ then she knows she has `prepared' Bob's
boxes in the state $\frac{1}{\sqrt{2}}\left(e^{-i\pi/8}\left|00\right\rangle +e^{i\pi/8}\left|11\right\rangle \right)$,
and she sends $c=0$. If she wishes to reveal $0$, she tells Bob
 she had input $0$ and obtained
$0$. If she wishes to reveal $1$, she tells Bob
she had input $1$ and obtained
$0$. Similarly, if she obtains $1$, she sends $c=1$, etc. It is straightforward to verify that this strategy gives rise to $P_{\mathrm{cont}}=\cos^{2}\left(\frac{\pi}{8}\right)\simeq 0.854$.

Using the GHZ state, dishonest Bob's strategy consists of having Alice measure $\sigma_y$ and $\sigma_x$ according to the value of her commitment. Bob then measures the polarization of one of his qubits along the $y$ axis and that of the other along the $x$ axis. Whenever his outcomes are correlated, in the event that Alice sends $c=0$ ($c=1$) he guesses that she has input $1$ ($0$), while whenever his outcomes are anti-correlated he guesses the reverse.
It is straightforward to verify that this strategy gives rise to an information gain of $\frac{3}{4}$.

{\em Device-independent coin flipping} -- 
(Strong) {coin flipping} is defined as the problem of two remote distrustful
parties having to agree on a bit. If both parties
are honest, then the outcome of the coin is uniformly random.  The degree of security afforded by a protocol is quantified
by the biases $\epsilon^A_{i}=P^A_{i}-\frac{1}{2}$ and $\epsilon^B_{i}=P^B_{i}-\frac{1}{2}$,
where $P^A_{i}$ ($P^B_{i}$) is Alice's (Bob's) maximal probability
of biasing the outcome to $i$. 
The quantity $\epsilon=\max\left\{ \epsilon^A_{i},\,\epsilon^B_{j}\right\}_{i,j}$ is usually referred to as the bias of the protocol. A protocol is said to be fair
whenever Alice and Bob enjoy the same bias.
Like {bit-commitment}, and indeed most non-trivial protocols
in distrustful cryptography, in the classical world its
security is completely breached if no limits are placed on a dishonest
party's computational power. In the quantum world the
story is different \cite{Aharonov}, the optimal bias is $\epsilon=0.207$ \cite{Kitaev, Chailloux and Kerenidis} (a weaker version of {coin flipping}, on the other hand, allows for arbitrarily small bias \cite{Mochon III}).

We remind the reader of a standard method to implement {coin flipping} using {bit-commitment}:  Alice commits to a random bit $a$, Bob sends a random bit $b$ to Alice, and then Alice reveals $a$. The outcome of the coin flip is just $a \oplus b$. In particular, $\epsilon^A_{i}=\epsilon_{\mathrm{cont}}$ and $\epsilon^B_{i}=\epsilon_{\mathrm{gain}}$. Using this construction with our {device-independent} {bit-commitment} protocol, we obtain a {device-independent} {coin flipping} protocol with biases $\epsilon^A_{i}=\cos^2\left(\frac{\pi}{8}\right)-\frac{1}{2}\simeq 0.354$ and $\epsilon^B_{j}=\frac{1}{4}$.

Since $\epsilon^A_{i}>\epsilon^B_{j}$, this construction advantages Alice. It is possible to lower the bias by equalizing the individual biases. Consider a new {coin flipping} protocol which consists of two repetitions of the above {coin flipping} protocol as follows. The result of the first (in which Alice commits) is used to determine who commits in the second. Say if the outcome is $0$ ($1$), then Alice (Bob) commits in the second. It is no longer a priori clear what strategy Alice should adopt in the first repetition, since, in principle, it may be beneficial for her to adopt one in which she sometimes loses the first coin flip, but increases her chances of making it to the second repetition (by not getting caught cheating in the first repetition in which case Bob aborts). Nevertheless, it is evident that Alice's maximal cheating probability is bounded from above by $\cos^4\left(\frac{\pi}{8}\right)+\left(1-\cos^2\left(\frac{\pi}{8}\right)\right)\cdot \frac{3}{4}\simeq{0.838}$. On the other hand, Bob never gets caught cheating in the first repetition (though he may of course lose), therefore Bob's maximal cheating probability is just $\frac{3}{4}\cos^2\left(\frac{\pi}{8}\right)+\frac{1}{4}\cdot\frac{3}{4} \simeq 0.827$. By allowing for more repetitions (the $n-1\,$th repetition determining who commits in the $n\,$th, etc.) we obtain that the biases $\epsilon^A_{i}$ and $\epsilon^B_{j}$ of the resulting protocol are bounded from above by  $\simeq 0.336$.

{\em Discussion} -- By introducing explicit {device-independent} {bit-commitment} and {coin flipping} protocols, we have shown that protocols in the distrustful cryptography model -- wherein Alice and Bob do not cooperate to estimate the amount of nonlocality present -- are amenable to a {device-independent} formulation. The fascinating connection between quantum nonlocality and cryptography, first noted by Ekert twenty years ago \cite{Ekert}, is thus seen to apply also in the very rich field of cryptography with mutually distrustful parties (and devices), affording us with a novel perspective on the connection between cryptography and the foundations of quantum mechanics.

To conclude, we would like to point out some notable features of our protocols. (i)  The protocols are single-shot and do not rely on any statistical estimation of the amount of nonlocality such as in the testing the degree of violation of a Bell inequality (even though their security is of course based on nonlocality). (ii) The {device-dependent} version of our protocol does not offer more security than the device-independent version. (iii) Since our security analysis is {device-independent}, it also covers the case where Alice's and Bob's outputs are affected by noise. Note that the analysis of noisy classical {coin flipping} in \cite{nguyen,hanggi} allows us to compute the quantum advantage in this case. (iv) The security afforded by our {device-independent} protocols is reasonably close to (though of course greater than) that of the best known {device-dependent} protocols. For the {bit-commitment} protocol we have $P_\mathrm{cont}\simeq 0.854$ and $P_\mathrm{gain}=\frac{3}{4}$, as compared to $P_\mathrm{cont}=P_\mathrm{gain}=\frac{3}{4}$ for the best known {device-dependent} protocol. The {coin flipping} protocol has a bias of $\lesssim 0.336$, as compared to $0.207$ in the {device-dependent} case. (v) Our work allows the study of {bit-commitment} and {coin flipping} in the context of theories other than quantum mechanics. Indeed, it relies only on the GHZ paradox (to define the protocol in the honest case), on  Tsirelson's bound on the CHSH inequality violation (which limits Alice's control) and on the no-signaling principle (which limits Bob's information gain). Curiously, the security of the protocol would increase if Tsirelson's bound were to decrease, reaching $P_{\mathrm{cont}}=P_{\mathrm{gain}}=\frac{3}{4}$ if it were equal to the local causal bound. In a theory constrained only by no-signaling, our protocol is no longer secure as PR boxes \cite{PR box} allow to maximally violate the CHSH inequality, implying $P_{\mathrm{cont}}=1$. Note that perfect {bit-commitment} was shown to be possible provided that honest parties have access to PR boxes and under the strong hypothesis (which we do not make) that a dishonest party cannot in any way tamper with the boxes \cite{Buhrman}. It is an open question whether there exists a quantum {bit-commitment}  protocol that is secure against dishonest parties limited only by the no-signaling principle, as is the case in quantum key-distribution~\cite{Barrett et al.,Masanes1}. 

{\em Acknowledgments} -- 
We acknowledge support from the BSF (grant no. 32/08) (N.A.), the
Inter-University Attraction Poles Programme (Belgian Science Policy) under Project IAP-P6/10 (Photonics@be) (S.M., S.P., J.S), a BB2B grant of the Brussels-Capital region (S.P.), the FNRS (J.S.), the  projects ANR-09-JCJC-0067-01, ANR-08-EMER-012 (A.C., I.K.), and the project QCS (grant 255961) of the E.U. (S.M., S.P., J.S., A.C., I.K.).

\end{document}